\input{aipcheck}

\documentclass[
    ,final            
  ]
  {aipproc}

\layoutstyle{6x9}

\newcommand{\fh}{f({\rm H_2})}

\newcommand{\ngrb}{11}

\newcommand{\lya}{Ly$\alpha$}

\def\h2{H$_2$}
\def\f0{$F_0$}

\newcommand{\cm}[1]{\, {\rm cm^{#1}}}

\newcommand{\sci}[1]{{\rm \; \times \; 10^{#1}}}
\newcommand{\mnhi}{N_{\rm HI}}

\newcommand{\nhi}{$N_{\rm HI}$}

\def\nhi{$N_{\rm HI}$}

\def\aap{A \& A}

\def\apj{ApJ}
\def\apjl{ApJL}

\def\apjs{ApJS}

\def\araa{ARAA}
\def\mnras{MNRAS}
\def\nat{Nature}

\begin{document}

\title{Resolving The ISM Surrounding GRBs
with Afterglow Spectroscopy}

\classification{<Replace this text with PACS numbers; choose from this list:
                \texttt{http://www.aip..org/pacs/index.html}>}

\author{J.X. Prochaska}{
address={Lick Observatory, University of California, Santa Cruz, CA 95064}
}

\author{H.-W. Chen}{
  address={Dept.\ of Astronomy \& Astrophysics and
Kavli Institute for Cosmological Physics,
5640 S.\ Ellis Ave, Chicago, IL, 60637, U.S.A.}
}

\author{M. Dessauges-Zavadsky}{
  address={Observatoire de Gen\`eve, 51 Ch. des 
      Maillettes, 1290 Sauverny, Switzerland }
}

\author{J. S. Bloom}{
  address={Department of Astronomy, 601 Campbell Hall, 
University of California, Berkeley, CA 94720-3411}
}

\begin{abstract}
We review current research related to spectroscopy 
of gamma-ray burst (GRB) afterglows with particular emphasis on 
the interstellar medium (ISM) of the galaxies hosting these 
high redshift events.
These studies reveal the physical conditions of 
star-forming galaxies and yield clues to the nature of the
GRB progenitor.
We offer a pedagogical review of the experimental design
and review current results.
The majority of sightlines are characterized by large HI column
densities, negligible molecular fraction, the ubiquitous detection
of UV pumped fine-structure transitions, and metallicities ranging
from 1/100 to nearly solar abundance.
\end{abstract}

\maketitle



\noindent {\bf Introduction:} 
With the discovery that long-duration ($t>2$s) gamma-ray bursts (GRBs)
are extragalactic events \cite{mdk+97}, it was immediately realized
that these phenomenon offer a means of 
probing gas in the early universe that complements spectroscopic
surveys of high $z$ quasars.  Identified exclusively with star-forming
galaxies \cite{bkd02,fls+06}, GRB events are believed to result
from the death of a massive star \cite{w93,wb06}.  
The majority emit a bright, power-law afterglow due to the deacceleration
of the relativistic jet in the surrounding 
interstellar or circumstellar medium.
The extraordinary luminosity of these events allows for
their detection and analysis at very high redshift
($z>10$), affording the opportunity to study the epoch of reionization
when bright quasars were extremely rare \cite{lr00}.
In addition to studies of the IGM, the GRB 
afterglow spectrum records data on the gas local to the event as
well as the interstellar medium surrounding the star-forming region.
In turn, analysis yields constraints on the physical conditions within
star-forming galaxies in the early universe and also clues to the
nature of the GRB phenomenon.

In this proceedings, I offer a pedagogical review of
current research using GRB afterglows to probe the universe.   
Much of the recent advances were enabled by the launch of the
{\it Swift} satellite, thanks to its rapid 
localization of GRB afterglows.
This proceeding describes the basics of research using
GRB afterglow spectroscopy and several of the key results 
related to the ISM of the host galaxies. 

\begin{figure}[ht]
\includegraphics[height=5.3in,angle=90]{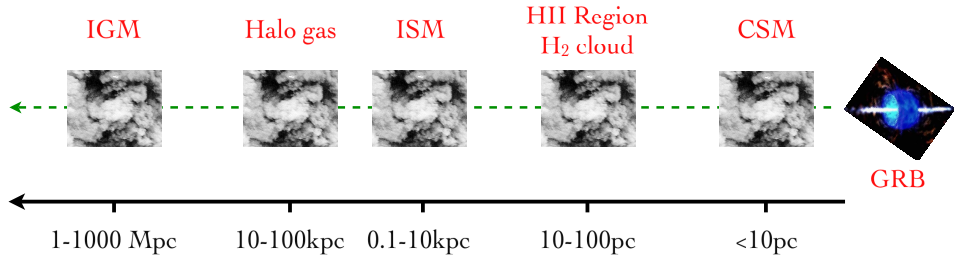}
\caption{Schematic of the various environments 
intersected by a GRB sightline.  Although these phases
may be well separated by distance, their absorption features
are only distinguished by differences in velocity.  Hubble
expansion separates the intergalactic medium (IGM), 
but the various phases within the host galaxy are likely to
overlap and one must focus on specific transitions to
isolate them.  
}
\label{fig:cartoon}
\end{figure}

\vskip 0.1in 

\noindent {\bf The Experiment:}
It is expected that GRBs occur within the star-forming
regions of high $z$ galaxies.  This assertion is supported
by the detection of star-forming galaxies coincident with the
GRB afterglow and the fact that GRBs tend
to occur in the bluest, brightest regions of these galaxies
\cite{bkd02,fls+06}.  The sightlines to the GRB afterglow,
therefore, will travel through 
(i)  the circumstellar medium surrounding the GRB progenitor;
(ii) the star-forming region hosting the GRB (e.g.\ a molecular 
cloud and/or HII region);
(iii) the ambient ISM of the host galaxy;
(iv) the baryonic halo of the galaxy;
and
(v) the intergalactic medium separating Earth from the GRB.

\begin{figure}[ht]
\includegraphics[height=5.3in,angle=90]{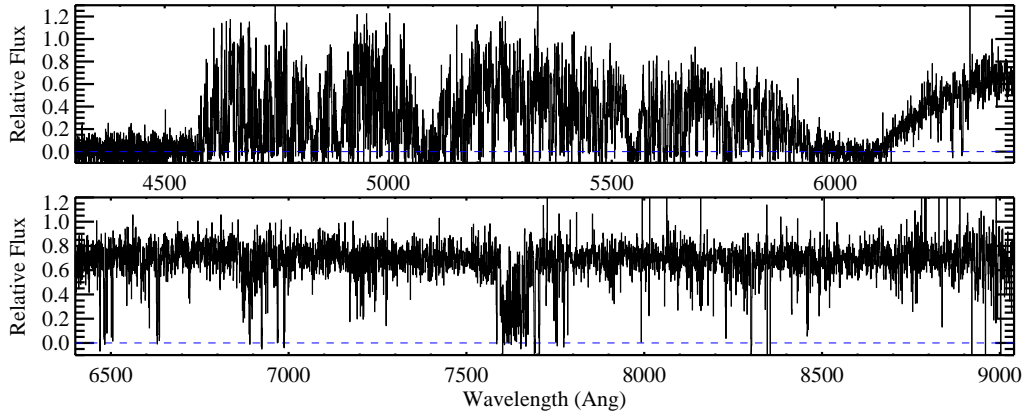}
\caption{MIKE/Magellan echelle spectrum of the bright
afterglow of GRB~050730 ($z_{GRB} = 3.96855$).  
Imposed on the
intrinsic power-law spectrum ($f_\nu \propto \nu^{-1.9}$)
are absorption features from the IGM at redshifts
$z<z_{GRB}$ and gas associated with the host galaxy of
GRB~050730.  These include strong Lyman series and
Lyman limit absorption by HI gas from the ambient ISM
(e.g.\ \lya\ at $\lambda \approx 6100$\AA)
and a series of strong metal-line transitions at wavelengths
$\lambda > 6200$\AA.
}
\label{fig:050730}
\end{figure}

Figure~\ref{fig:cartoon} presents a cartoon describing this
experiment and the various phases along the sightline.  
The gas in these various regions
will imprint signatures on the observed power-law afterglow
spectrum.  Ideally, one would separate these spectral features
according to the distance of the gas from the GRB to isolate 
analysis on specific regions (e.g.\ the circumstellar material).
Unfortunately, spectra only distinguish the velocity of gas
and one is forced to disentangle these various phases.  
This is relatively straightforward for the IGM; Hubble
expansion gives a lower redshift to all of the gas foreground
to the GRB host galaxy.  The regions within the host galaxy,
however, may share similar velocities and 
one must identify unique signatures (i.e.\ specific spectral
transitions) to disentangle their imprints on the afterglow spectrum.

In Figure~\ref{fig:050730} we present the afterglow spectrum
of GRB~050730 obtained with the MIKE echelle spectrometer
on the Magellan~I telescope \cite{cpb+05}.  Redward of 
6400\AA\ one observes the intrinsic power-law spectrum 
of the afterglow due to synchrotron radiation.
The absorption features imposed on this power-law spectrum are
associated with gas foreground to the afterglow.  
The thicket of absorption lines at $\lambda < 6000$\AA\ are
the IGM, the so-called \lya\ forest.
Similar to the featureless blazars, 
the IGM reveals the redshift of the GRB, $z_{GRB}$, i.e.,  
the \lya\ forest terminates at the redshift of the GRB host galaxy.  
From the spectrum
we estimate $z_{GRB} \approx 6100{\rm \AA}/1215.67{\rm \AA} - 1 \approx 4.0$.
The following sections describe a number of scientific results
related to studies of spectra like these.

\vskip 0.1in

\noindent {\bf Hydrogen Gas:}
The strongest features in the spectrum of GRB~050730 (Figure~\ref{fig:050730})
lie at $\lambda < 4600$\AA\ and $\lambda \approx 6100$\AA. 
These correspond to the Lyman
limit and \lya\ profiles of the HI gas of the GRB host galaxy. 
The spectrum easily resolves the 
Lorentzian damping wings of the \lya\ profile which give this
spectral feature its name: a damped \lya\ (DLA) system.  
A Voigt profile fit to the line-profile yields a precise
measurement of the HI column density, e.g.\  $\mnhi = 10^{22.15} \cm{-2}$.


Large HI column densities are a common feature in GRB afterglow
spectra \cite{vel+04,jfl+06}; GRB sightlines tend
to intersect large surface densities of interstellar medium.
The median \nhi value is $\approx 10^{21.7} \cm{-2}$ and 90\%\
exceed $10^{20} \cm{-2}$ although there is a tail that extends
down to $10^{17} \cm{-2}$.
In contrast to the GRB sightlines, out of the $\approx 1000$
quasar sightlines exhibiting damped \lya\ absorption
($\mnhi \ge 2 \sci{20} \cm{-2}$), only a few have $\mnhi > 10^{21.7} \cm{-2}$
and none exceed $10^{22} \cm{-2}$.
This difference emphasizes 
how GRB sightlines complement
quasars observations: GRB sightlines probe the tiny 
regions with high surface density that are associated
with star-forming regions in high $z$ galaxies.

The large HI column densities of GRB sightlines were initially 
interpreted as the signatures of molecular clouds \cite{rp02}.
The column density distribution, however, includes too many 
sightlines with $\mnhi < 10^{21} \cm{-2}$ to be 
uniquely described by this model \cite{jfl+06}.  Furthermore,
as we describe below, the majority of this neutral gas
has negligible molecular fraction and 
is located at $r > 100$\,pc from the afterglow.  
This distance
exceeds all but the very largest molecular clouds in local
galaxies.  Current expectation, therefore, is that the \nhi\ values
reflect the ambient ISM of the host galaxy \cite{pcd+07}.  

With sufficient blue coverage and resolution,
the spectra enable measurements of the surface density
of molecular hydrogen H$_2$ along GRB sightlines \cite{vel+04,tpc+07}.
Surprisingly, the molecular fractions are very low: $\fh < 10^{-6}$.
There are two important implications of these results.
First, it demonstrates that the molecular cloud which
(presumably) hosted the GRB was photoevaporated, at least along
the sightline.  The absence of excited H$_2$ states in the spectrum
indicates the cloud was destroyed prior to the GRB event.
Second, H$_2$ formation is being suppressed throughout the
GRB host galaxy.  This implies an intense far-UV radiation field,
roughly $100\times$ the Galactic value \cite{tpc+07},    
presumably produced by star-formation throughout
the galaxy.  This hypothesis
should be tested with deep, multi-band images of the GRB
host galaxies.

\begin{figure}[ht]
\includegraphics[height=5.3in,angle=90]{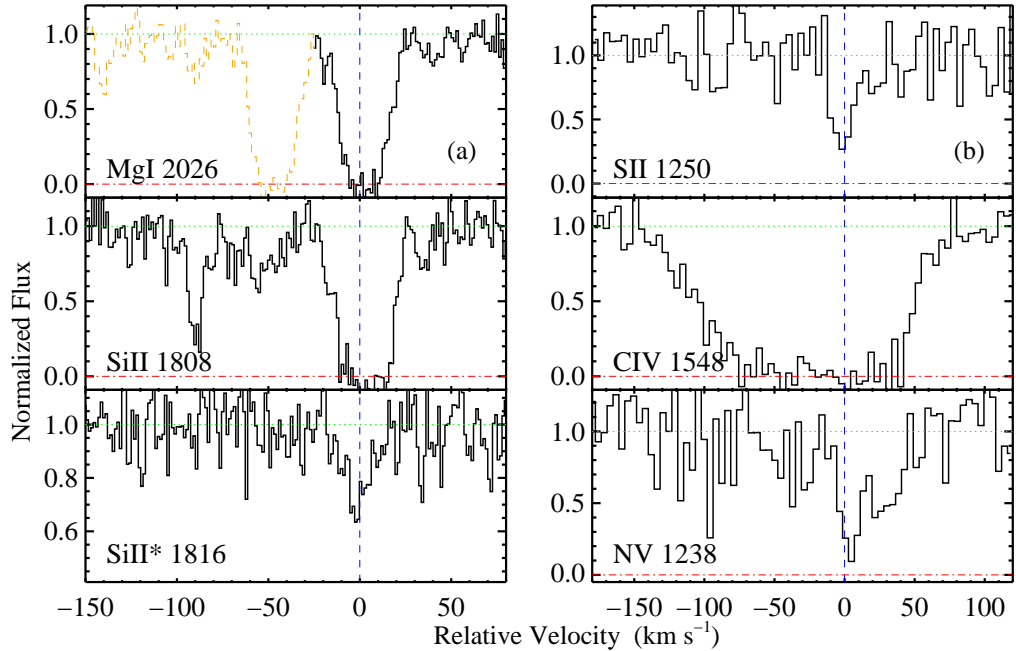}
\caption{ Diversity of metal-line transitions identified with the host
galaxies of (a) GRB~051111 and (b) GRB~050730 \cite{pcb+07,cpb+05}.
The MgI and fine-structure (SiII*)
transitions constrain the majority of neutral gas to occur within
100\,pc to a few kpc from the GRB afterglow.  The resonance, low-ion
transitions give the metallicity and abundance pattern of
the ambient ISM.  Finally, the high-ion transitions (CIV, NV) 
trace the galactic halo and gas near ($r \approx 10$\,pc)
the GRB progenitor.
}
\label{fig:metals}
\end{figure}

\vskip 0.1in

\noindent {\bf Metal-Line Transitions:}
In addition to the Lyman series, the GRB spectra 
generally show metal-line transitions with
large equivalent width \cite{bsc+03,sff03}
reflecting the very large HI surface
densities typical of the sightlines.
Figure~\ref{fig:metals} presents the diverse set of 
metal-line transitions observed for two GRB sightlines.
These include transitions from atomic states with 
ionization potential (IP~$< 1$\,Ryd; MgI),
resonance and fine-structure transitions from low-ion
states (1\,Ryd$ < $IP$< \approx 2$\,Ryd; FeII, SiII*), resonance transitions
from high-ion transitions (2\,Ryd$<$IP$< 4$\,Ryd; SiIV, CIV),
and resonance transitions from extremely high-ions 
(IP$>4$\,Ryd; NV).
This diversity of ionic transitions indicates the sightlines
penetrate several phases associated with the host 
galaxy (Figure~\ref{fig:cartoon}).  

\vskip 0.05in

\noindent {\it Resolving the Distance of the Gas:}
The neutral gas observed along GRB sightlines almost
universally shows two sets of transitions that allow one
to constrain its distance from the GRB afterglow: MgI and
the fine-structure transitions of Fe$^+$, Si$^+$, and O$^0$.
The detection of the former, which coincides in velocity
with the low-ion transitions (i.e.\ bulk of the neutral gas), 
sets a lower limit of the gas at $\approx 100$\,pc.
If this MgI gas were closer to the afterglow, it 
would have been ionized prior to the spectral observations
\cite{Mir03,pcb06}.  

While the detection of MgI sets a lower limit to the
distance of the neutral gas from the GRB afterglow, the
ubiquitous detection \citep{bsc+03,vel+04,cpb+05,bpck+05}
of fine-structure transitions 
(e.g.\ SiII* 1264) places an upper limit to the 
distance \cite{pcb06}.  
For a full discussion of the astrophysics of these
transitions,  see the contribution by P. Vreeswijk.
The principle conclusion is that 
in the GRB environment (i.e.\ near the extraordinarily bright
afterglow) indirect UV pumping dominates the excitation \cite{pcb06}.
This has been confirmed by observations of 
variability in the populations of the fine-structure 
levels \cite{dcp+06,vls+07}.  As the afterglow fades, the UV excitation
rate decreases and the majority of excited levels will depopulate.
Detailed analysis indicates the gas lies at 100\,pc to 2\,kpc.

There are a number of implications and applications associated
with the UV pumping of fine-structure transitions \cite{pcb06}.
First, information related to the density and temperature
of the gas has been washed out by the UV pumping.  
Second, a precise accounting of the chemical abundances of the
gas must include these fine-structure levels ($\approx 20\%$ correction).  
Third, the detection of fine-structure transitions
sets an upper limit on the distance to the gas of a few kpc \cite{pcb06}.
Fourth, if one observes strong resonance-line absorption
without corresponding fine-structure absorption the gas must be located
at distances greater than a few kpc.
This has allowed observers to demonstrate that absorption lines previously
interpreted as circumstellar material are in fact at much greater
distance from the GRB event \cite{cpr+07}.  It has also been
valuable for resolving the origin of low-ion velocity fields \cite{pcw+08}.
Finally, one concludes that the majority of neutral gas along the GRB
sightline is located at 100\,pc to a few kpc from the afterglow, i.e.\  
beyond the immediate star-forming region 
of the GRB progenitor.
While these data do 
not directly constrain the environment of the GRB progenitor,   
we recall that much of
our knowledge on the progenitor comes from global studies of the
host galaxy (e.g.\ redshift, star-formation history, morphology).

\vskip 0.05in

\noindent {\it Metallicity:}
Significant attention has been given to the 
metallicity of the gas in galaxies hosting GRBs.  This 
has been driven by theoretical expectation that the GRB progenitor
should have low metallicity to minimize mass-loss during the
lifetime of the star. If the star loses too
much mass, it will likely have too little angular momentum to 
drive the relativistic jet which powers the GRB and its afterglow.
Current theory suggests that the GRB progenitors will have
metallicities less than 1/10 solar abundance under the assumption
of the single-star, collapsar model \cite{wh06,ln06}. 

\begin{figure}[ht]
\includegraphics[height=5.0in,angle=90]{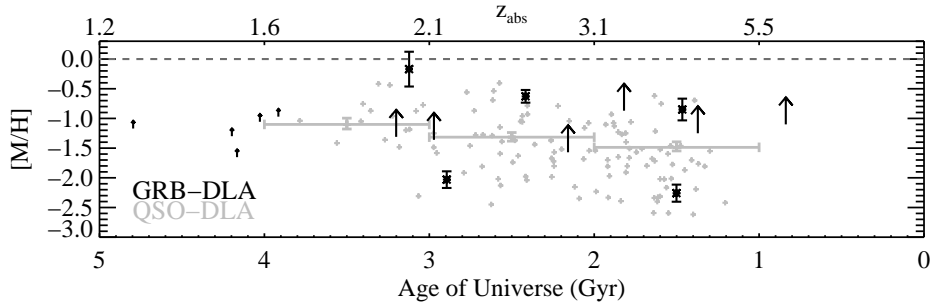}
\caption{ Metallicity distribution of the GRB population 
(dark blue) as a function of redshift.  The sample is restricted
to $z>2$ where one measures the metallicity from the GRB afterglow
spectrum, i.e.\ a comparison of the metal column density derived
from low-ion transitions against the HI column density measured
from the \lya\ profile \cite{pcd+07}.  
The GRB distribution spans two decades in metallicity and has
a median value that likely exceeds 1/10 solar.  
The green points indicate the
metallicities of galaxies (DLAs) intersecting high $z$
quasar sightlines \cite{pgw+03}.  These also span a wide range
of metallicity but have systematically lower abundance than
the GRB sample.  
}
\label{fig:mh}
\end{figure}

At low redshift ($z_{GRB}<1$), there is mounting evidence that GRBs preferentially
occur in metal-poor galaxies.   Imaging of the host galaxies reveal
low luminosities and irregular morphologies characteristics of 
low metallicity galaxies in the local universe \cite{cvf+05,fls+06}.
More directly,  emission-line diagnostics provide estimates of the
metallicity in the star-forming regions of these galaxies 
indicating sub-solar abundances but 
rarely less than 1/10 solar \cite{pbc+04,sof+05,mha+06}. 

At high redshift ($z_{GRB} > 2$), one can measure the metallicity distribution 
of gas near the GRB using afterglow spectroscopy.
With absorption-line techniques one simply counts atoms,
avoiding the many uncertainties associated with calibrating emission-line
diagnostics.  There are, however, a few potential sources of error:
ionization corrections, dust depletion, and line-saturation.
The first issue is minor because of the very large HI column densities;
the majority of gas will be in the low-ion states, i.e.\ 
Si$^+$/H$^0$ $\approx$ Si/H.
Depletion effects are minimized by
evaluating the metallicity with non-refractory or mildly
refractory elements (e.g.\ S, O, Zn).
The final issue is problematic \cite{pro06}; 
with lower resolution or S/N data one is generally
only able to report a lower limit to the metallicity.

In Figure~\ref{fig:mh} we present the metallicities of \ngrb\ 
GRBs with $\mnhi > 10^{20} \cm{-2}$ \cite{pcd+07}.  The sample exhibits
a wide distribution of metallicities ranging from 1/100 solar to 
nearly solar metallicity.  Presumably this diversity reflects the
diversity of the underlying galaxy population hosting GRBs at these
redshifts.   Importantly, the distribution is consistent with 
drawing galaxies randomly according to their current
SFR at $z \approx 3$ \cite{fps+08}. 
The figure also indicates that the GRB metallicity distribution
has a median value of at least 1/10 solar;
a strict metallicity upper limit is not supported by
the empirical observations.

It is also illuminating to compare the GRB metallicity distribution
with a control sample.  In Figure~\ref{fig:mh} we present the metallicity
values of $\approx 100$ DLAs intervening high $z$ quasars \cite{pgw+03}.
These galaxies show a similar range of metallicity as the GRB sample
yet the distribution is systematically lower \cite{bpck+05,pcd+07}.
This is likely because QSO-DLA (selected by HI cross-section, not
current star-formation) correspond to the outer regions of somewhat
fainter galaxies \cite{fps+08}.

\begin{theacknowledgments}
J. X. P. is partially supported by NASA/Swift grants 
NNG06GJ07G and NNX07AE94G and an NSF CAREER grant (AST-0548180).
\end{theacknowledgments}






\end{document}